# KELT: The Kilodegree Extremely Little Telescope, a Survey for Exoplanets Transiting Bright, Hot Stars

Joshua Pepper
Department of Physics, Lehigh University, Bethlehem, PA, United States
e-mail: joshua.pepper@lehigh.edu

Keivan Stassun
Department of Physics and Astronomy, Vanderbilt University, Nashville, TN, United States

B. Scott Gaudi
Department of Astronomy, The Ohio State University, Columbus, OH, United States

**Abstract**
The KELT project was originally designed as a small-aperture, wide-field photometric survey that would be optimally sensitive to planets transiting bright (V~8-10) stars. This magnitude range corresponded to the gap between the faint magnitude limit where radial velocity surveys were complete, and the bright magnitude limit for transiting planet hosts routinely found by dedicated ground-based transit surveys. Malmquist bias and other factors have also led the KELT survey to focus on discovering planets transiting relatively hot host stars as well. To date, the survey has discovered 22 transiting hot Jupiters, including some of the brightest transiting planet host stars known to date. Over half of these planets transit rapidly-rotating stars with $T_{eff} > 6250$ K, which had been largely eschewed by both radial velocity and transit surveys, due to the challenge of obtaining precision radial velocities for such stars. The KELT survey has developed a protocol and specialized software for confirming transiting planets around stars rotating as rapidly as ~200 km/s. This chapter reviews KELT planet discoveries, describes their scientific value, and also briefly discusses the non-exoplanet science produced by the KELT project, especially long-timescale phenomena and preparations for the TESS mission.

## Introduction

In the early 2000's the field of exoplanet discovery was mostly limited to radial-velocity (RV) surveys. By 2000 there were 31 exoplanets discovered, all using the RV method, except for the three pulsar planets discovered by pulsar timing (Wolszczan & Frail 1992). (All historical information related to exoplanet discovery here is based on information in the NASA Exoplanet Archive (https://exoplanetarchive.ipac.caltech.edu/) queried in November 2017.) Astronomers were actively using photometric follow-up to determine whether any of these RV-discovered planets transited their host stars. Simultaneously, there were several surveys that were trying to use the transit method to discover planets. The RV-discovered "Hot Jupiter" planet HD 209458b was the first exoplanet observed to transit its host star (Charbonneau et al 2000; Henry et al 2000), thereby conclusively demonstrating that "Hot Jupiters" were indeed objects of planetary mass, radius, and density. However, it was not until 2003 that the first exoplanet was discovered with the transit method, OGLE-TR-56b (Udalski et al. 2002, 2003; Konacki et al 2003).

During that time, a large number of efforts were being undertaken to discover exoplanets with the transit method. Two survey strategies were generally employed. One type of search used existing telescopes in campaigns lasting for a few days or weeks. Such projects included the EXPLORE and EXPLORE/OC surveys (Mallen-Ornelas et al 2003; von Braun et al 2003), PISCES (Mochejska et al 2002), STEPSS (Burke et al 2003), OGLE (Udalski et al 1992), among others. Those surveys observed relatively small fields of a few square degrees intensively over the course of several days to a few weeks, mostly observing stellar clusters or parts of the Galactic plane. The other strategy involved the use of dedicated, small-aperture robotic telescopes observing large regions of the sky (thousands of square degrees) over the course of months or years. One of the first of these was the Vulcan survey (Borucki et al 2001), which was shortly followed by WASP/SuperWASP (Pollacco et al 2006), TrES (Alonso et al 2004), HATNet (Bakos et al 2004), and XO (McCullough et al 2005), and later QES (Alsubai et al 2013).

At that same time, Pepper et al (2003) published the first comprehensive theoretical examination of the optimal configuration of a small-aperture wide-field telescope for transit detection. That paper prompted the construction of the Kilodegree Extremely Little Telescope (KELT; Pepper et al 2004). The KELT telescope (later renamed KELT-North) was deployed to Winer Observatory in Arizona in 2004, with regular survey operations beginning in 2005 (Pepper et al 2007). After Vanderbilt University joined the project, a near-duplicate of that telescope was constructed to observe the southern hemisphere. That telescope, KELT-South (Pepper et al 2012), was deployed to the South African Astronomical Observatory (SAAO) observing site in Sutherland, South Africa in 2008, with regular survey operations beginning in 2010.

Since that time, Ohio State, Vanderbilt, and Lehigh Universities have jointly operated the KELT survey, with both telescopes regularly observing the sky, building up a time-series photometric catalogue of several million bright stars. The KELT project has discovered nearly two dozen exoplanets, and has yielded a number of additional scientific publications on other topics. This chapter summarizes the scientific goals of the KELT project, the technical configuration of the KELT telescopes, the primary challenges of the effort, and the resulting discoveries.

## Scientific Goals and the KELT Telescopes

KELT was designed from the beginning with the goal of discovering planets transiting bright host stars. While the meaning of the term "bright" in this context can vary, for the design of KELT it refers to the magnitude regime just fainter than the magnitude where RV surveys were complete, roughly $V \approx 8$ (Wright et al. 2011). Planets transiting such bright host stars are intrinsically valuable, as they offer the greatest opportunity for detailed follow-up investigation. Pepper, Gould, & DePoy (2003), determined the optimal design for a single telescope with a goal of detecting transits of stars with $V \approx 8 - 10$ primarily by defining the required aperture and field of view needed to achieve that goal.

The KELT telescopes each use a Mamiya 645 80 mm f/1.9 medium-format manual-focus lens with a 42 mm aperture. They each have a $4096 \times 4096$ thermoelectrically-cooled

Apogee camera (AP16E for KELT-North and Alta U16M for KELT-South). This lens and camera combination provides a roughly 23 arcsec/pixel image scale and a 26° × 26° field of view. The large field of view is crucial to the ability of KELT to survey a large fraction of the sky with a single camera.

Each clear night, the KELT telescopes observe a pre-selected set of fields. Each telescope rotates among all fields that are above an airmass of 1.5, using an exposure time of 150 seconds. The field locations are shown in Figure 1, along with a possible distribution of fields from the upcoming TESS mission. There are 44 fields observed by KELT-North, and 26 fields observed by KELT-South, and altogether cover about 70% of the sky. There is a small amount of overlap between the two telescopes near the celestial equator. The KELT field locations were chosen through a set of considerations including observability, scientific interest, synergies with other surveys, and both positive and negative implications of observing near the Galactic plane and the ecliptic.

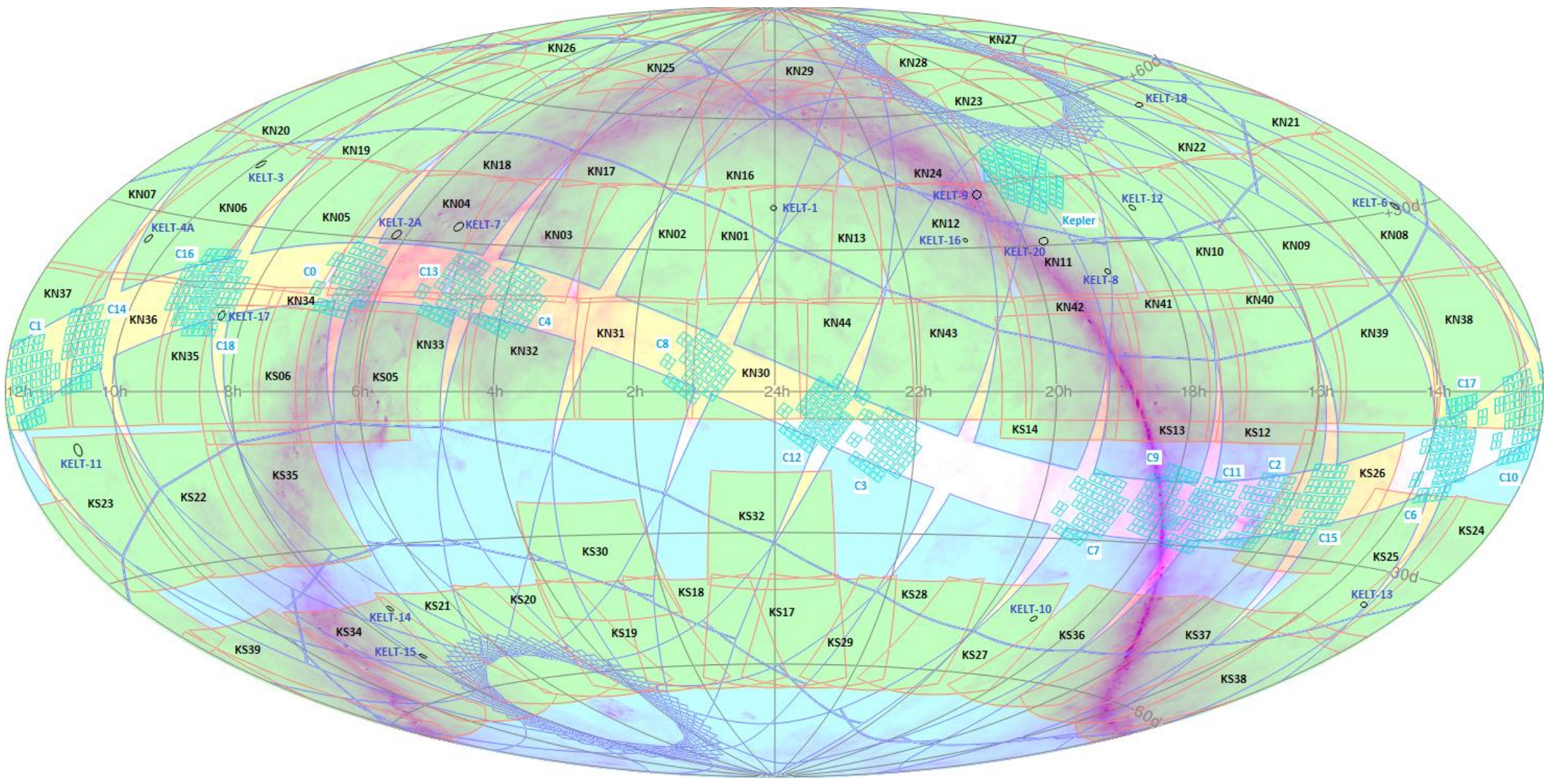

Figure 1: Locations of the KELT fields (green), the Kepler and K2 fields (cyan), and a sample location layout for the TESS fields (blue). The Galaxy is indicated in magenta, and the locations of the published KELT planets are also indicated.

Images are reduced and photometry extracted using a set of procedures based on customized versions of the ISIS difference imagining software package (Alard & Lupton 1998, Alard 2000). Siverd et al (2012) provides a comprehensive review of the details of the reduction and candidate selection criteria, including the use of the BLS (Kovacs et al 2002) and TFA (Kovacs et al 2005) algorithms developed by members of the HAT project team. It is worthwhile noting that the candidate selection criteria outlined in Siverd et al (2012) were developed before any candidates were identified and followed up. These criteria were defined by injecting fake transit signals into actual KELT light curves, and then using the BLS algorithm to determine what fraction of those injected signals were reliably recovered. In this way, the KELT project was able to optimize the ratio of real signals to false positives, as well as automate the candidate selection procedure to the largest extent possible. Once candidates are selected, a series of observational tests are conducted to confirm their planetary nature, or to identify false positives. Additional details of the follow-up and confirmation process are described in Collins et al (in prep).

## The KELT Exoplanet Discoveries

To date, KELT has discovered 22 transiting exoplanets. Twenty of them have been published, and the remaining two are in the process of final analysis and paper submission. Table 1 lists the published KELT exoplanets and their key properties. The primary goal of the KELT survey was to discover transiting exoplanets orbiting bright stars. That goal has been accomplished, with the discovery of two planets brighter than $V = 8$, and nine planets brighter than $V = 10$. Indeed, the two brightest known hosts of a transiting Hot Jupiter (in the $V$ band) were discovered by KELT (with one of these, KELT-20, simultaneously discovered by the MASCARA collaboration). Figure 2 shows distribution of the KELT exoplanets, compared to the full set of known transiting exoplanets, in optical magnitude and effective temperature space.

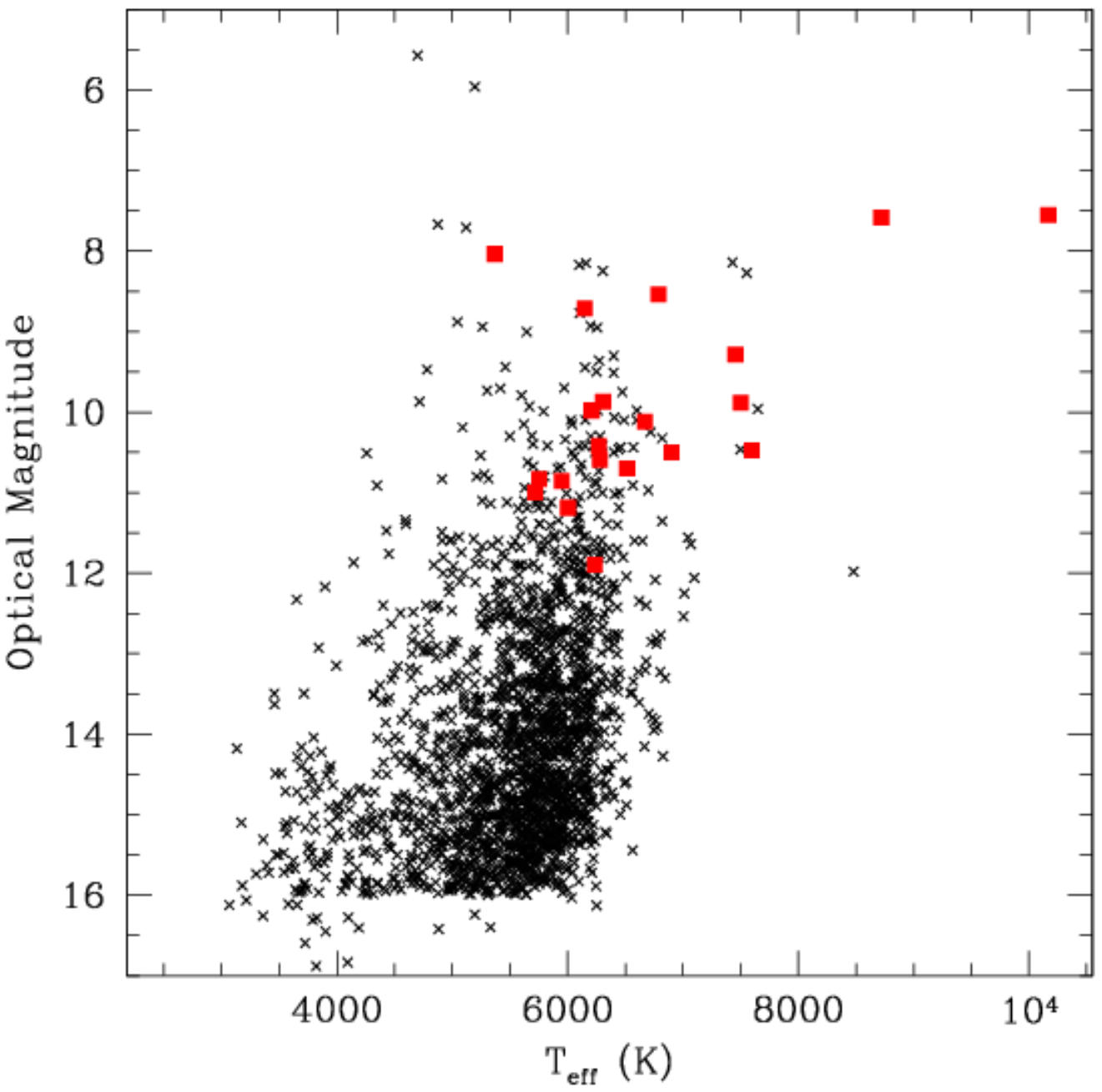

Figure 2: Distribution of transiting planet hosts by optical magnitude and $T_{eff}$. Black crosses are known transit hosts, red boxes are the KELT transit hosts.

An additional feature of the KELT survey design that was not appreciated during the first years of operation was the degree to which a magnitude-limited survey such as KELT was most sensitive to hot and evolved stars. That is due to normal Malmquist bias, and the nature of the stellar population in the $7 < V < 11$ magnitude range to which KELT is sensitive. The hosts of KELT transit candidates are therefore generally hotter than for most other planet surveys (Bieryla et al 2015). Also, since A and early F stars are above the Kraft break (Kraft 1967), they tend to be fast rotators. This rapid rotation can make measurement of the radial velocity reflex motion of the host star due to the transiting planet difficult or impossible to measure, thereby making the 'traditional' method of confirming a transiting planet also difficult or impossible.

However, as demonstrated in Collier Cameron et al. 2010 and codified in Bieryla et al 2015, it is nevertheless possible to confirm transiting planets around rapidly-rotating hot stars. The basic steps are as follows. One first confirms that the transit signal is real via higher-precision follow-up photometry. Second, one uses relatively imprecise (~100 m/s) radial velocities to obtain an upper limit on the stellar reflex motion to confirm that, if the occulting body indeed orbits the primary, it must have a planetary mass. Finally one confirms that the occulting body indeed orbits the primary by measuring the Doppler Tomography (DT) signal of the planet during the transit.

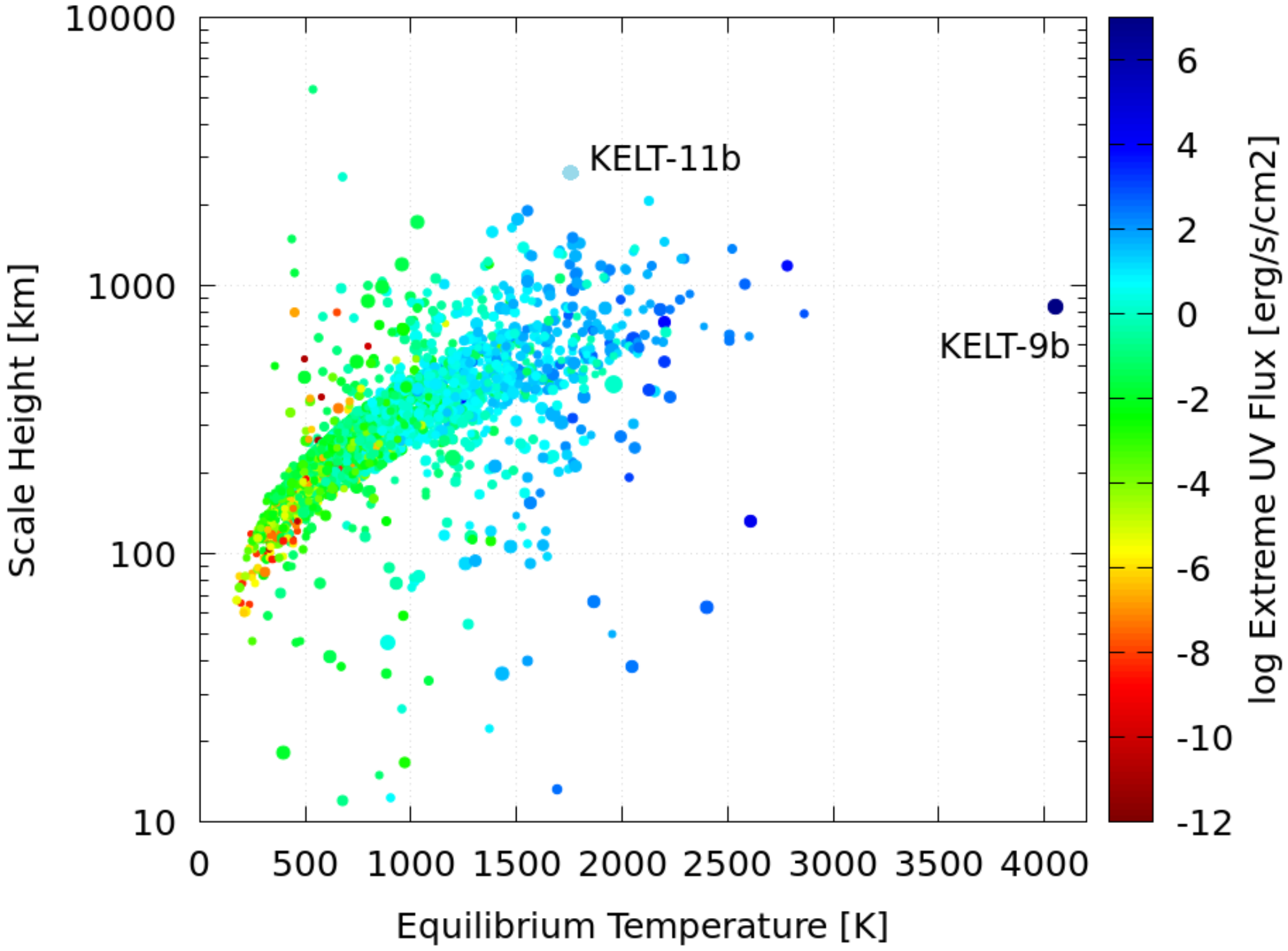

Figure 3: Atmospheric scale height versus equilibrium temperature for known transiting planets with measured masses. Colour (key at right) represents the amount of incident extreme-ultraviolet ($\lambda \leq 91.2$ nm) flux the planet receives from its parent star, and the symbol size is inversely proportional to the V magnitude of the host. KELT-9b is hotter than any other known transiting gas giant planet by approximately 1,000 K and receives about 700 times more extreme-ultraviolet flux. KELT-11b has one of the largest estimated scale heights of any known planet.

Although this sounds straightforward in practice, the combination of stellar temperatures and high stellar rotation requires careful use and analysis of RV follow-up methods. Most of the KELT planet discoveries could not have been made without access to the TRES spectrograph and pipeline, which has been optimized to be able to extract RV information from hot and fast-rotating stars (Zhou et al 2016). The key component of this method of confirming transiting planets, DT, was first demonstrated by Collier Cameron et al (2007). Indeed, the DT technique has been used to confirm several of our planets (KELT-17b: Zhou et al 2016; KELT-9b: Gaudi et al 2017; KELT-13b/WASP-167b: Temple et al 2017; KELT-20b/MASCARA-2b: Lund et al 2017; KELT19Ab: Siverd et al 2018; KELT-21b: Johnson et al 2018).

As described in detail in the introduction to the discovery of KELT-19Ab (Siverd et al 2018), a combination of a few coincidental and nearly simultaneous occurrences, as well as the ‘late entry’ of KELT into the field of exoplanet discovery via transits, led the project to focus on the discovery of transiting planets around hotter stars.

Many of the KELT discoveries are noteworthy for unique or extreme properties. A few of the most significant are:

- KELT-1b (Siverd et al 2012): This discovery was and remains the brightest host star with a transiting brown dwarf, and is one of the best-characterized brown dwarfs known (Beatty et al 2017).
- KELT-9b (Gaudi et al 2017): This is a giant planet transiting a 10,170 K star straddling the B and A spectral types, with V=7.55. The planet has an equilibrium temperature of 4600K, making is as hot as a mid-K star. Any molecules in the atmosphere of this planet are expected to be fully dissociated. The incident UV flux is orders of magnitude greater than any other known planet, and thus the planet should be experiencing significant mass loss due to evaporation.
- KELT-11b (Pepper et al 2017): KELT-11b is one of the most inflated planets known, with a sub-Saturn mass but a radius 40% larger than Jupiter. Its estimated atmospheric scale height is over 2700 km, and is the brightest transit host star in the southern hemisphere.

The distribution of known transiting planets in terms of host star temperature and the planet atmospheric scale height is shown in Figure 3, with the noteworthy locations of KELT-9 and KELT-11 indicated.

**Table 1: Properties of the Published KELT Exoplanets and Host Stars**

| Name | V-mag | Period (days) | Host $T_{eff}$ (K) | Host Mass ($M_\odot$) | Host Radius ($R_\odot$) | Planet Mass ($M_J$) | Planet Radius ($R_J$) | $T_{eq}$ (K) | K-mag | Transit Depth |
|---|---|---|---|---|---|---|---|---|---|---|
| KELT-1b | 10.7 | 1.2175 | 6516 | 1.34 | 1.47 | 27.38 | 1.12 | 2423 | 9.44 | 0.0061 |
| KELT-2Ab | 8.77 | 4.1138 | 6148 | 1.31 | 1.84 | 1.524 | 1.29 | 1714 | 7.35 | 0.0053 |
| KELT-3b | 9.8 | 2.7034 | 6306 | 1.28 | 1.47 | 1.477 | 1.35 | 1811 | 8.66 | 0.0089 |
| KELT-4Ab | 9.98 | 2.9896 | 6206 | 1.20 | 1.60 | 0.902 | 1.70 | 1827 | 8.69 | 0.012 |
| KELT-6b | 10.38 | 7.8456 | 6102 | 1.09 | 1.58 | 0.43 | 1.19 | 1313 | 9.08 | 0.0060 |
| KELT-7b | 8.54 | 2.7348 | 6789 | 1.54 | 1.73 | 1.28 | 1.53 | 2048 | 7.54 | 0.0083 |
| KELT-8b | 10.8 | 3.2441 | 5754 | 1.21 | 1.67 | 0.867 | 1.86 | 1675 | 9.18 | 0.013 |
| KELT-9b | 7.55 | 1.4811 | 10170 | 2.52 | 2.36 | 2.88 | 1.89 | 4050 | 7.48 | 0.0068 |
| KELT-10b | 10.7 | 4.1662 | 5948 | 1.11 | 1.21 | 0.679 | 1.40 | 1377 | 9.34 | 0.014 |
| KELT-11b | 8.03 | 4.7365 | 5370 | 1.44 | 2.72 | 0.195 | 1.37 | 1712 | 6.12 | 0.0027 |
| KELT-12b | 10.64 | 5.0316 | 6279 | 1.59 | 2.37 | 0.95 | 1.78 | 1800 | 9.36 | 0.0060 |
| KELT-13b / WASP 167b | 10.5 | 2.0220 | 6900 | 1.59 | 1.79 | <8 | 1.58 | 2329 | 9.76 | 0.0082 |
| KELT-14b / WASP-122b | 11.0 | 1.7101 | 5802 | 1.18 | 1.37 | 1.196 | 1.52 | 1904 | 9.42 | 0.013 |
| KELT-15b | 11.2 | 3.3294 | 6003 | 1.18 | 1.48 | 0.91 | 1.44 | 1642 | 9.85 | 0.01 |
| KELT-16b | 11.7 | 0.9690 | 6236 | 1.21 | 1.36 | 2.75 | 1.42 | 2453 | 10.64 | 0.011 |
| KELT-17b | 9.23 | 3.0802 | 7454 | 1.64 | 1.65 | 1.31 | 1.53 | 2087 | 8.65 | 0.0091 |
| KELT-18b | 10.1 | 2.8718 | 6670 | 1.52 | 1.91 | 1.18 | 1.57 | 2120 | 9.21 | 0.0072 |
| KELT-19Ab | 9.89 | 4.6117 | 7500 | 1.62 | 1.83 | <4.07 | 1.91 | 1935 | 9.20 | 0.011 |
| KELT-20b / MASCARA-2b | 7.6 | 3.4741 | 8730 | 1.76 | 1.56 | <3.52 | 1.74 | 2261 | 7.42 | 0.013 |
| KELT-21b | 10.48 | 3.6128 | 7598 | 1.46 | 1.64 | <3.91 | 1.59 | 2051 | 10.09 | 0.0099 |

## The KELT Light Curve Catalog and Survey Legacy

The KELT photometric data have been used for a large set of astrophysical investigations beyond exoplanet science. These include the study of young stars with circumstellar disks, in which the warped or disrupted disks eclipse their own host star (Rodriguez et al 2013, Rodriguez et al 2016a, Rodriguez et al 2016b, Kennedy et al 2017a, Rodriguez et al 2017b, Osborn et al 2017). This project also discovered the longest-period eclipsing binary star ever found (Rodriguez, et al. 2016c), along with a set of stars in young clusters showing odd variability patterns, which are of interest for observations by K2 and TESS (Rodriguez, et al. 2017a; Ansdell, et al. 2018). Other results include studies of Be stars (Labadie-Bartz et al 2017, 2018), supernovae (Siverd et al 2015, Goobar et al 2015), and stellar rotation (Cargile et al 2014, Terrien et al 2014).

There are roughly 5 million stars with light curves from KELT. Part of the data set has been released through the NASA Exoplanet Archive, and plans are underway for the release of the remainder. With the upcoming TESS mission poised to detect nearly all transiting exoplanets orbiting bright stars on short-period orbits, the long-term value of the KELT observations rests in the unique combination of its photometric properties. That includes the reasonably high photometric precision, sky coverage, magnitude range, and time baseline. An initial characterization of the variability of stars observed by KELT has been published (Oelkers et al 2018), and more detailed investigations of particular classes of variables are underway.

Another ancillary benefit of the KELT project has been the creation of the KELT-Follow-Up Network (KFUN; Collins et al in prep). Because KELT did not have access to sufficient observing resources for follow-up and confirmation of the resulting transit candidates, the KELT project created a collaboration of professionals and amateur observers to assist with that process. KFUN includes mostly amateur astronomers and professional astronomers from small colleges with access to telescopes that are typically 0.3m to 0.7m in aperture. They are well-suited for photometric follow-up of the bright KELT transit candidates. This process has been enhanced by the development within the KELT collaboration of the TAPIR software for scheduling transit observations (Jensen 2013) and the AstroImageJ package for acquiring and analysing time-series photometry (Collins & Kielkopf 2013, Collins et al 2017).

The KFUN collaboration now includes over 90 separate institutions and individual partners across the world, across much of North America, Europe, and Australia. In the coming era of TESS and LSST, there will be an enormous need for follow-up observations of variable and transient events discovered by surveys, across all areas of astrophysics. The KFUN collaboration will continue to operate to assist with that effort after the KELT survey itself concludes, and will represent a significant legacy from that project.

**Acknowledgements**
We would like to thank the members of the KELT science team for making this project possible, including Thomas Beatty, Karen Collins, Knicole Colon, Jason Eastman, David James, Marshall Johnson, Rudolf Kuhn, Jonathan Labadie-Bartz, Michael Lund, Ryan Oelkers, Joseph Rodriguez, Robert Siverd, Daniel Stevens, Xinyu Yao, and George Zhou. We would especially like to thank Xinyu Yao for assistance in preparing this manuscript. We would also like to thank all the members of the KELT Follow-Up Network for long-term support of the KELT project.

## References

Alard C (2000) Image subtraction using a space-varying kernel. A&AS, 144, 363

Alard C & Lupton RH (1998) A Method for Optimal Image Subtraction. ApJ, 503, 325

Alonso R, Brown TM, Torres G et al (2004) TrES-1: The Transiting Planet of a Bright K0 V Star. ApJL, 613, L153

Alsubai KA, Parley NR, Bramich DM et al (2013) The Qatar Exoplanet Survey. ACTAA, 63, 465

Ansdell M, Oelkers RJ, Rodriguez JE et al. (2018) Identification of young stellar variables with KELT for K2 - II. The Upper Scorpius association. MNRAS, 473, 1231

Bakos GA, Noyes RW, Kovacs G et al (2004) Wide-Field Millimagnitude Photometry with the HAT: A Tool for Extrasolar Planet Detection. PASP, 116, 266

Beatty TG, Madhusudhan N, Pogge R et al (2017) The Broadband and Spectrally Resolved H-band Eclipse of KELT-1b and the Role of Surface Gravity in Stratospheric Inversions in Hot Jupiters. AJ, 154, 242

Bieryla A, Collins K, Beatty TG et al (2015) KELT-7b: A Hot Jupiter Transiting a Bright V = 8.54 Rapidly Rotating F-star. AJ, 150, 12

Borucki WJ, Caldwell D, Koch DG et al (2001) The Vulcan Photometer: A Dedicated Photometer for Extrasolar Planet Searches. PASP, 113, 439

Burke CJ, Depoy DL, Gaudi BS, Marshall JL (2003) Survey for Transiting Extrasolar Planets in Stellar Systems (STEPSS): The Frequency of Planets in NGC 1245. Scientific Frontiers in Research on Extrasolar Planets, 294, 379

Cargile PA, James DJ, Pepper J et al (2014) Evaluating Gyrochronology on the Zero-age-main-sequence: Rotation Periods in the Southern Open Cluster Blanco 1 from the KELT-South Survey. ApJ, 782, 29

Charbonneau D, Brown TM, Latham DW, Mayor M (2000) Detection of Planetary Transits Across a Sun-like Star. ApJ, 529, 45

Collier Cameron A, Guenther E, Smalley B et al (2010) Line-profile tomography of exoplanet transits – II. A gas-giant planet transiting a rapidly rotating A5 star. MNRAS, 407, 507

Collins K & Kielkopf J (2013) AstroImageJ: ImageJ for Astronomy. Astrophysics Source Code Library, record ascl:1309.001

Collins KA, Kielkopf JF, Stassun KG, Hessman FV (2017) AstroImageJ: Image Processing and Photometric Extraction for Ultra-precise Astronomical Light Curves. AJ, 153, 77

Gaudi BS, Stassun KG, Collins KA et al (2017) A giant planet undergoing extreme-ultraviolet irradiation by its hot massive-star host. Nature, 546, 514

Goobar A, Kromer M, Siverd R et al (2015) Constraints on the Origin of the First Light from SN 2014J. ApJ, 799, 106

Henry GW, Marcy GW, Butler RP, Vogt SS (2000) A Transiting ``51 Peg-like" Planet. ApJ, 529, 41

Jensen E (2013) Tapir: A web interface for transit/eclipse observability. Astrophysics Source Code Library, record ascl:1306.007

Johnson, MC, Rodriguez JE, Zhou G et al (2018) KELT-21b: A Hot Jupiter Transiting the Rapidly-Rotating Metal-Poor Late-A Primary of a Likely Hierarchical Triple System. AJ, 155, 100

Kennedy GM, Kenworthy MA, Pepper J et al (2017) The transiting dust clumps in the evolved disc of the Sun-like UXor RZ Psc. Royal Society Open Science, 4, 160652

Konacki M, Torres G, Jha S, Sasselov DD. (2003) An extrasolar planet that transits the disk of its parent star. Nature, 421, 507

Kovacs G, Zucker S, Mazeh T (2002) A box-fitting algorithm in the search for periodic transits. A&A, 391, 369

Kovacs G, Bakos G, Noyes RW (2005) A trend filtering algorithm for wide-field variability surveys. MNRAS, 356, 557

Kraft RP (1967) Studies of Stellar Rotation. V. The Dependence of Rotation on Age among Solar-Type Stars. ApJ, 150, 551

Labadie-Bartz J, Pepper J, McSwain MV et al (2017) Photometric Variability of the Be Star Population. AJ, 153, 252

Labadie-Bartz J, Chojnowski SD, Whelan DG et al (2018) Outbursts and Disk Variability in Be Stars. AJ, 155, 53

Lund MB, Rodriguez JE, Zhou G et al (2017) KELT-20b: A Giant Planet with a Period of P~3.5 days Transiting the V~7.6 Early A Star HD 185603. AJ, 154, 194

Mallen-Ornelas G, Seager S, Yee HKC et al (2003) The EXPLORE Project. I. A Deep Search for Transiting

Extrasolar Planets. ApJ, 582, 1123
McCullough PR, Stys JE, Valenti JA et al (2005) The XO Project: Searching for Transiting Extrasolar Planet Candidates. PASP, 117, 783
Mochejska BJ, Stanek KZ, Sasselov DD, Szentgyorgyi AH, et al (2002) Planets in Stellar Clusters Extensive Search. I. Discovery of 47 Low-Amplitude Variables in the Metal-rich Cluster NGC 6791 with Millimagnitude Image Subtraction Photometry. AJ, 123, 3460
Oelkers RJ, Rodriguez JE, Stassun KG et al (2018) Variability Properties of Four Million Sources in the TESS Input Catalog Observed with the Kilodegree Extremely Little Telescope Survey. AJ, 155, 39
Osborn HP, Rodriguez JE, Kenworthy MA et al (2017) Periodic eclipses of the young star PDS 110 discovered with WASP and KELT photometry. MNRAS, 471, 740
Pepper J, Gould A, Depoy DL (2003) Using All-Sky Surveys to Find Planetary Transits. ACTAA, 53, 213
Pepper J, Gould A, Depoy DL (2004) KELT: The Kilodegree Extremely Little Telescope. The Search for Other Worlds, 713, 185
Pepper J, Pogge R., Depoy DL et al (2007) Early Results from the KELT Transit Survey. Transiting Extrapolar Planets Workshop ASP Conference Series, Vol. 366, p.27
Pepper J, Kuhn RB, Siverd R, James D, Stassun K (2012) The KELT-South Telescope. PASP, 124, 230
Pepper J, Rodriguez JE, Collins KA et al (2017) KELT-11b: A Highly Inflated Sub-Saturn Exoplanet Transiting the V = 8 Subgiant HD 93396. AJ, 153, 215
Pollacco DL, Skillen I, Collier Cameron A et al (2006) The WASP Project and the SuperWASP Cameras. PASP, 118, 1407
Rodriguez JE, Pepper J, Stassun K et al (2013) Occultation of the T Tauri Star RW Aurigae A by its Tidally Disrupted Disk. AJ, 146, 112
Rodriguez JE, Reed PA, Siverd RJ et al (2016a) Recurring Occultations of RW Aurigae by Coagulated Dust in the Tidally Disrupted Circumstellar Disk. AJ, 151, 29
Rodriguez JE, Stassun KG, Cargile P et al (2016b) DM Ori: A Young Star Occulted by a Disturbance in Its Protoplanetary Disk. ApJ, 831, 74
Rodriguez JE, Stassun KG, Lund MB et al (2016c) An Extreme Analogue of ε Aurigae: An M-giant Eclipsed Every 69 Years by a Large Opaque Disk Surrounding a Small Hot Source. AJ, 151, 123
Rodriguez JE, Ansdell M, Oelkers RJ et al (2017a) Identification of Young Stellar Variables with KELT for K2. I. Taurus Dippers and Rotators. ApJ, 848, 97
Rodriguez JE, Zhou G, Cargile PA et al (2017b) The Mysterious Dimmings of the T Tauri Star V1334 Tau. ApJ, 836, 209
Siverd RJ, Beatty TG, Pepper J et al (2012) KELT-1b: A Strongly Irradiated, Highly Inflated, Short Period, 27 Jupiter-mass Companion Transiting a Mid-F Star. ApJ, 761, 123
Siverd RJ, Goobar A, Stassun KG, Pepper J (2015) Observations of the M82 SN 2014J with the Kilodegree Extremely Little Telescope. ApJ, 799, 105
Temple LY, Hellier C, Albrow MD et al (2017) WASP-167b/KELT-13b: joint discovery of a hot Jupiter transiting a rapidly rotating F1V star. MNRAS, 471, 2743
Terrien RC, Mahadevan S, Deshpande R et al (2014) New Red Jewels in Coma Berenices. ApJ, 782, 61
Udalski A, Szymanski M, Kaluzny J, Kubiak M, Mateo, M (1992) The Optical Gravitational Lensing Experiment. ACTAA, 42, 253
Udalski A, Paczynski,B., Zebrun K., Szymanski M et al (2002) The Optical Gravitational Lensing Experiment. Search for Planetary and Low-Luminosity Object Transits in the Galactic Disk. Results of 2001 Campaign. AcA, 52, 1
Udalski A, Zebrun K, Szymanski M, Kubiak M, et al (2002) The Optical Gravitational Lensing Experiment. Search for Planetary and Low- Luminosity Object Transits in the Galactic Disk. Results of 2001 Campaign – Supplement. AcA, 42, 115
von Braun K, Lee B, Mallen-Ornelas G et al (2003) EXPLORE/OC: A Search for Planetary Transits in the Field of NGC 2660. ArXiv:astro-ph/0312217
Wolszczan A, Frail DA (1992) A planetary system around the millisecond pulsar PSR1257 + 12. Nature, 355, 145
Wright JT, Marcy GW, Howard AW, Johnson, JA et al. The Frequency of Hot Jupiters Orbiting nearby Solar-type Stars. PASP, 123, 412
Zhou G, Rodriguez JE, Collins KA et al (2016) KELT-17b: A Hot-Jupiter Transiting an A-star in a Misaligned Orbit Detected with Doppler Tomography. AJ, 152, 136